\documentclass[a4paper,10pt]{article}
\usepackage[utf8]{inputenc}
\usepackage{amsmath}

\title{ Supersymmetry Breaking as a  new source for  the Generalized Uncertainty Principle }
\author{  Mir Faizal  \\
\\ Deptartment of Physics and Astronomy, University of Lethbridge, \\  Lethbridge, AB T1K 3M4, Canada}
\date{}
\begin{document}

\maketitle

\begin{abstract}
In this letter,  we will demonstrate that the    breaking of supersymmetry  by a
non-anticommutative deformation can be used to 
  generate the generalized uncertainty principle.  
We will  analyse  the physical reasons for this observation, in the framework of string theory. 
We also discuss  the relation between the generalized uncertainty principle and the Lee-Wick  field theories. 
\end{abstract}
As  supersymmetry has not been observed, it is expected that the supersymmetry has to be broken as sufficiently large energy scales, 
and various mechanisms have been  proposed for  breaking of the supersymmetry at such scales \cite{4}. A partial breaking of supersymmetry can also occur 
due to a non-anticommutative deformation of the original theory  \cite{5}-\cite{5d}. 
The non-anticommutative deformation of a theory has been motivated from the  noncommutative deformation of ordinary field theories 
\cite{2}-\cite{2c},  where spacetime coordinates do not commute. As supersymmetric theories can be analysed in superspace 
(which contains additional coordinates with odd Grassman parity), 
it is possible to impose non-anticommutativity on these coordinates with odd Grassman parity. In this case, the 
ordinary products of fields is replaced by a non-anticommutativity Moyal product of fields. 
It is also possible to analyse a total breaking 
of supersymmetry by imposing a different kind of non-anticommtativity \cite{6}. In this paper, we will demonstrate that this  
non-anticommutative deformation   will generate the  generalized uncertainty principle for 
theories where all the supersymmetry is broken. 

It may be noted that the generalized uncertainty principle has been originally motivated by the existence of a minimum measurable length 
scale in nature \cite{g}-\cite{g1}. The existence of such 
a minimum measurable length can be proven from the  physics of black holes. This is because  the 
energy required to probe a region of space below Planck scale is less than the energy required to form a mini black hole in that region 
of space \cite{z4}-\cite{z5}.  So, any attempt to probe a phenomena below Planck scale will lead to the formation of a mini black hole in that 
region of space, and this will prevent any measurement in that region. Furthermore,  
almost all approaches to quantum gravity predict the  existence of such  a minimum measurable length in spacetime. 
The string theory is one of the most interesting approaches to quantum gravity, and strings are smallest probe  
in perturbative string theory. So,  it is not possible to probe spacetime below string length scale.  Hence, string length 
acts as a minimum measurable length in string theory \cite{a}.  Even in loop quantum gravity, it is the existence of a minimum measurable length 
which   turns 
the big bang into a big bounce \cite{z1}.  However,   the existence of such a minimum measurable length scale is not consistent with 
the usual Heisenberg uncertainty principle, and so the  usual Heisenberg uncertainty principle has to be  generalized 
  to a generalized uncertainty principle to incorporate the existence of such a minimum measurable length  
  \cite{g}-\cite{g1}. The    deformation of the Heisenberg uncertainty principle also leads to a deformation 
  of the Heisenberg algebra,  and this in turn deforms the 
coordinate representation of the momentum operator. 
 A covariant formalism of the generalized uncertainty principle has been used to deform the equations of motion of quantum field theory, 
  and the action for such deformed theories has also been studied \cite{skdj}-\cite{h}.   In this paper, 
  we will demonstrate that this  deformation  produced from the 
  generalized uncertainty principle is exactly the same as the deformation which breaks all the  supersymmetry of a supersymmetric  
  field theory. Hence, the breaking of supersymmetry  
  can be used as another motivation to study generalized uncertainty principle. 
  
 A four dimensional supersymmetric field theory with $\mathcal{N} = 1$ supersymmetry is parameterized 
 as $(x^\mu, \theta^{\alpha},\bar{\theta}^{\dot{\alpha}})$, where  $x^{\mu}$
is a spacetime vectors, and  $\theta^{\alpha}$,
$\bar{\theta}^{\dot{\alpha}}$  are the two component Weyl spinors. A free Wess-Zumino
model can be defined on this  superspace as 
\begin{eqnarray}
S & = & \frac{1}{2}\int d^4 x d^2  \theta d^2 \bar \theta [\Phi \bar{\Phi}+\bar{\Phi} \Phi].
\end{eqnarray}
In terms of component fields, this action can be written as 
\begin{eqnarray}
S &=&\int d^{4}x[A\partial^\mu \partial_\mu \bar{A}+i\partial_{\mu}\psi\sigma^{\mu}\bar{\psi}  +F\bar{F}].
\end{eqnarray}
Thus, usual non-anticommutative deformation of the supersymmetric field theory occurs by deforming $\theta^\alpha$  as 
$\left\{ \theta^{\alpha},\theta^{\beta}\right\} = C^{\alpha\beta}\thinspace,\label{eq:1}
$
where $C^{\alpha\beta}$ is a two-dimensional  matrix which causes the deformation of the theory \cite{5}-\cite{5d}. 
Here $\bar{\theta}^{\dot{\alpha}}$ is not deformed. The supersymmetric field theory constructed on this superspace also 
gets deformed by this deformation of the superspace. In  this deformed supersymmetric field theory  all the product of fields are replaced  by
the Moyal star product of those fields. 
However, recently a different kind of deformation of non-anticommutative field theories has been constructed, 
and  this deformation is defined as \cite{6}
\begin{eqnarray}
\left\{ \theta^{\alpha},\bar{\theta}^{\dot{\alpha}}\right\} _{\star} & = &\xi C^{\alpha\dot{\alpha}}\thinspace.\label{Moyalthetas}
\end{eqnarray}
Here $\xi C^{\alpha\dot{\alpha}}$ is again a two-dimensional  matrix which causes this deformation, and 
$\xi$ is related to scale at which supersymmetry will be broken.  This deformation breaks all the supersymmetry of a 
four dimensional supersymmetric field theory with $\mathcal{N} = 1$ supersymmetry. 
It is useful to define $|C|=\frac{1}{2}C^{\alpha\dot{\alpha}}C^{\beta\dot{\beta}}\epsilon_{\alpha\beta}\epsilon_{\dot{\alpha}\dot{\beta}}$
as the determinant of $C^{\alpha\dot{\alpha}}$. 
It is also possible to define a Molyar star product corresponding to this deformation as,  
\begin{eqnarray}
f \star g  & = &f \exp\left[\frac{\xi}{2}C^{\alpha\dot{\alpha}}\left(\stackrel{\leftarrow}{D}_{\alpha}
\stackrel{\rightarrow}{\bar{D}}_{\dot{\alpha}}+\stackrel{\leftarrow}{\bar{D}}_{\dot{\alpha}}\stackrel{\rightarrow}{D}_{\alpha}\right)\right]g \nonumber \\
 & =&fg+\frac{\xi}{2}(-1)^{s_{f}}\left[\left(D_{\alpha}f\right)\left(\bar{D}_{\dot{\alpha}}g\right)+\left(\bar{D}_{\dot{\alpha}}f\right)\left(D_{\alpha}g\right)\right]\nonumber \\
 && -\frac{\xi^{2}}{16}|C|\left[\left(D^{2}f\right)\left(\bar{D}^{2}g\right)+\left(\bar{D}^{2}f\right)\left(D^{2}g\right)\right]\nonumber \\
 && -\frac{\xi^{2}}{8}C^{\alpha\dot{\alpha}}C^{\beta\dot{\beta}}\left[(\bar{D}_{\dot{\beta}}D_{\alpha}f)(\bar{D}_{\dot{\alpha}}D_{\beta}g)\right.\nonumber \\
 && \left.\hspace{1cm}+(D_{\beta}\bar{D}_{\dot{\alpha}}f)(D_{\alpha}\bar{D}_{\dot{\beta}}g)\right]. \label{Moyal}
\end{eqnarray}
The deformation of the   free Wess-Zumino
model  can be written as  
\begin{eqnarray}
S & = & \frac{1}{2}\int d^4 x d^2  \theta d^2 \bar \theta  [\Phi\star\bar{\Phi}+\bar{\Phi}\star\Phi]. 
\end{eqnarray} 
Now  in the component form, the deformation of the free Wess-Zumino model can be expressed as \cite{6}
\begin{eqnarray}
S & = &\int d^{4}x\left[A(1-\xi^{2}|C|\partial^\nu \partial_\nu )\partial^\mu \partial_\mu \bar{A}+
i\partial_{\mu}\psi\sigma^{\mu}(1-\xi^{2}|C|\partial^\nu \partial_\nu )\bar{\psi}\right.\nonumber \\
 & & \left. \hspace{1cm}+F(1-\xi^{2}|C|\partial^\nu \partial_\nu )\bar{F} \right]\ .\label{actionBoxcomponents}
\end{eqnarray}
This is because  $\Phi\star\bar{\Phi}  = \Phi\bar{\Phi}+\frac{\xi}{2}C^{\alpha\dot{\alpha}}(D_{\alpha}\Phi)(\bar{D}_{\dot{\alpha}}
\bar{\Phi})  -\frac{\xi^{2}}{16}|C|(D^{2}\Phi)(\bar{D}^{2}\bar{\Phi})\ , $ and 
$ \bar{\Phi}\star\Phi =  \bar{\Phi}\Phi+\frac{\xi}{2}C^{\alpha\dot{\alpha}}(\bar{D}_{\dot{\alpha}}\bar{\Phi})(D_{\alpha}\Phi) 
  -\frac{\xi^{2}}{16}|C|(\bar{D}^{2}\bar{\Phi})(D^{2}\Phi) \ $.\label{barPhiPhi}
  
We will demonstrate that the fermionic part of this deformed action will be exactly the same as the fermionic action deformed
by the   generalized uncertainty principle.
We can write the fermionic part of this deformed action as
\begin{eqnarray}
S & =&
i\partial_{\mu}\psi\sigma^{\mu}(1-\xi^{2}|C|\partial^\nu \partial_\nu )\bar{\psi} \nonumber \\
&=& 
i\partial_{\mu}\psi\sigma^{\mu}(1-\beta \partial^\nu \partial_\nu )\bar{\psi}, 
\end{eqnarray}
where  we have define  a new parameter   $\beta$ as 
\begin{eqnarray}
\beta    =  \xi^2 |C|
\end{eqnarray}
The equation of motion for this fermion field is given by 
\begin{eqnarray}
 i \gamma^\mu \partial_\mu (1 - \beta \partial^\nu \partial_\nu) \psi &=&0. 
\end{eqnarray}
It might be noted that this equation can also be produce by substituting 
\begin{eqnarray}
p_\mu = \tilde p_\mu (1  + \beta  \tilde p^\nu \tilde p_\nu)
\end{eqnarray}
in the following equation $
\gamma^\mu p_\mu \psi=0,  
$
where $ \tilde p_\mu = -i \partial_\mu $.
Thus, we can write the coordinate representation of this modified momentum operator as 
\begin{eqnarray}
 p_\mu  &=& -i \partial_\mu (1  - \beta   \partial^\nu \partial_\nu).
\end{eqnarray}
The coordinate representation of this modified operator is 
 exactly equal deformation of the coordinate representation of 
the momentum operator is obtained by the following deformation of the Heisenberg algebra
\begin{equation}
 [p^\mu, x_\nu] = i \delta^\mu_\nu + i \beta (\delta^\mu_\nu p^\tau p_\tau + 2 p^\mu p_\nu). 
\end{equation}
This deformation of the Heisenberg algebra occurs due to a deformation of the usual uncertainty principle 
to a generalized uncertainty principle \cite{d}-\cite{h}. In fact, this algebra  \cite{d}-\cite{h} is a covariant version of the usual 
deformed by the generalized uncertainty principle   \cite{g}-\cite{g1}, and it corresponds to the existence of both a minimum measurable  length 
and a minimum measurable time. This is because the   generalized uncertainty  corresponding to this deformed algebra is \cite{d}
\begin{eqnarray} 
\Delta x^\mu \Delta p_\mu &\geq&  \frac{1}{2}\left(1+\beta \Delta p^\rho \Delta p_\rho
+\beta \langle p^\rho \rangle \langle p_\rho \rangle\right)+i\left(\beta \Delta p^\mu
\Delta p_\mu+\beta\langle p^\mu \rangle \langle p_\mu \rangle\right)
\nonumber \\ &=&\frac{1}{2}\left(1+3\beta \Delta p^\mu \Delta p_\mu
+3\beta \langle p^\rho \rangle \langle p_\rho \rangle\right).
\end{eqnarray}
This in turn can be used to argue for the existence of a minimum measurable length $l_s$ and a minimum measurable time $t_s$, 
in the theory, where
\begin{eqnarray}
 l_s &= \sqrt{3\beta} &= \sqrt{3\xi^2 |C|}, \nonumber \\
 t_s &= \sqrt{3\beta} &= \sqrt{3 \xi^2 |C|}. 
\end{eqnarray}
It is possible that for non-relativistic systems, the  
the temporal part of such a deformation might be neglected \cite{m}, and in this case, we can write \cite{g}-\cite{g1}
\begin{eqnarray}
 [p^i, x_j] = i \delta^i_j + i \beta (\delta^i_j p^k p_k + 2 p^i p_j)
\end{eqnarray}
The coordinate representation of the momentum operator can now be written as 
\begin{eqnarray}
 p_i =  -i (1 - \beta \partial^k \partial_k )\partial_i 
\end{eqnarray} 
The     generalized uncertainty principle corresponding to this deformed Heisenberg algebra,
for the simple one dimensional case, can be written as 
\begin{eqnarray}
 \Delta x \Delta \geq \frac{1}{2} (1 + \beta (\Delta p)^2). 
\end{eqnarray}
Thus, we can argue that the generalized uncertainty principle   occurs because of the breaking  of   supersymmetry.

It may be noted that the relation between the non-anticommutativity 
and generalized uncertainty principle is something that was expected to occur. This is because the non-anticommutativity occurs due to background fluxes
in string theory as a $\alpha'$ level effect as $ C^{\alpha \beta } = \alpha' F^{\alpha \beta},$ and similarly $ C^{\alpha \dot{\alpha} } 
= \alpha' F^{\alpha \dot{\alpha} }$. Furthermore, as string have an extended structure, and they are also the smallest probes that can be used 
in string theory, it is not possible to probe below the string length scale. Thus, the string length $\sqrt{4\pi \alpha'}$ will act as the minimum 
measurable length scale, deforming the Heisenberg algebra to $ 
{\left[x^\mu, p_\nu \right] } = i \delta^\mu_\nu  + i  f( 4\pi  \alpha' ) \left[ \delta^\mu_\nu p^\tau p_\tau  + 2   p^\mu  p_\nu \right] $, where
$f( \alpha')  \to 0$, if strings are considered a point i.e.,   $\alpha' \to 0$ \cite{faiz}. This will deform the 
coordinate representation of the momentum operator as 
$ \partial_\mu \to    (1 - f( 4\pi { \alpha'} ) \partial^\tau \partial_\tau )\partial_\mu $. Now the leading order behavior  of $f(4\pi \alpha')$ is 
$f(\alpha')\sim \alpha'$ and the leading order behavior of  $\beta$ is 
$\beta = \xi^2 |C| \sim \alpha' $ as $C^{\alpha \dot{\alpha}} = \alpha' F^{\alpha \dot{\alpha}}$.
Thus, both the deformation are 
of $\alpha'$ order, and this seems to be the reason that the non-anticommutativity can be related to the 
generalized uncertainty principle. However, we have explicitly demonstrated that the non-anticommutative deformation 
of the  free Wess-Zumino model produces exactly the same results as the results obtained from the generalized uncertainty principle. 
Hence, it can be used as a new motivation to study the  generalized uncertainty principle.

It may   be noted that the non-anticommutativity will produce higher derivative terms, and so it will be 
 problematic to view it as a fundamental theory. However, we view it as an effective field theory description
of some more fundamental theory, where scales greater than a certain scale have been integrated out. We would like to point out 
that even though higher derivative terms have several problems associated with them, they are still being studied as they 
occur in various different approaches to quantum gravity. In this paper, we do not claim to resolve these problems 
associated with these higher derivative terms, but only demonstrate that the form of the higher derivative terms that occurs due 
to the deformation of a field theory by non-anticommutativity is exactly the same as the higher derivative terms produced by the generalized 
uncertainty principle. We would also like to point out that the
the generalized uncertainty principle
occurs in almost all theories of quantum gravity. So, non-anticommutativity should not be taken 
as the main reason for the occurrence of generalized uncertainty principle. It should rather be taken as another motivation for 
 the occurrence of generalized uncertainty principle.

Finally, we will like to comment that 
the field theory obtained from non-anticommutative deformation \cite{6}, and generalized uncertainty principle \cite{d}, resembles a
Lee-Wick  field theory \cite{w1}-\cite{w2}. The Lee-Wick  extension of the standard model has also been constructed \cite{stan}. 
It was observed that this Lee-Wick extension of the standard model 
stabilizes the Higgs mass against quadratically divergent radiative corrections. 
It has been demonstrated that the   in the Lee-Wick extension of the standard model,  the familiar see-saw mechanism for 
generating neutrino masses preserves the solution to the hierarchy puzzle provided by the higher derivative terms \cite{stand}. 
The flavor changing neutral currents have also been studied in the Lee-Wick extension of the standard model \cite{stand1}. 
As the    generalized uncertainty principle can be used to obtain a Lee-Wick extension of the fermionic fields, it
would be interesting to investigate  the consiquences of this relation between 
this Lee-Wick extension model and  the generalized uncertainty principle. 

\section*{Acknowledgments}
We would like to thank Saurya Das for useful discussions.


\begin{thebibliography}{99}
\bibitem{4} S. J. Gates Jr , M. T. Grisaru,  M. Rocek and  W.  Siegel,  Front. Phys. 58, 1 (1983)


\bibitem{5}   N.  Berkovits, N.  Seiberg,   JHEP 0307, 010 (2003) 
\bibitem{5a}
   A. Borowiec, J. Lukierski, M. Mozrzymas, V. N. Tolstoy,  JHEP 1206, 154  (2012)
   \bibitem{5b}  O.  Lunin, Soo-Jong Rey, JHEP 0309, 045  (2003)  
   \bibitem{5c}
A.  F. Ferrari, M. Gomes, J. R. Nascimento, A. Y. Petrov and A. J. da Silva,
  Phys. Rev.D  74, 125016 (2006)
  \bibitem{5d} M. Faizal, D. J. Smith, Phys. Rev. D87, 025019 (2013) 

  

\bibitem{2}  N. Seiberg and E. Witten, JHEP 09, 032 (1999)
\bibitem{2a}J. W. Moffat, Phys. Lett. B491, 345 (2000)
\bibitem{2b} E. Witten, Nucl. Phys. B460 , 335 (1996) 
\bibitem{2c}A. Astashkevich, N. Nekrasov and A. Schwarz, Commun. Math. Phys. 211, 167 (2000)
  
  \bibitem{6}   M. Dias, A. F. Ferrari, C. A. Palechor, C. R. Senise Jr, J. Phys. A 48,   275403  (2015)

  
  
  
\bibitem{g}   L. J. Garay, Int. J. Mod. Phys. A 10, 145  (1995)  

\bibitem{7q} S. Das, and E. C. Vagenas,  Phys. Rev. Lett.   101,  221301 (2008) 
\bibitem{0000a1} M. Liu, Y. Gui and H. Liu,  Phys. Rev. D78, 124003 (2008) 
\bibitem{0000a2} L. N.  Chang, Z.  Lewis, D.  Minic and T.  Takeuchi, Adv. High Energy Phys.  2011, 493514 (2011) 
\bibitem{greene}
  R. Easther, B. R. Greene, W. H. Kinney and G. Shiu,
  Phys.  Rev.  D {  67}, 063508 (2003)

 \bibitem{c}A. F. Ali, S. Das and E. C. Vagenas, Phys. Rev. D   84 , 044013 (2011)

\bibitem{g1}M. Maggiore, Phys. Lett. B 304,   65 (1993)

\bibitem{z4} M. Maggiore, Phys. Lett. B304, 65 (1993)
\bibitem{z5}M. I. Park, Phys. Lett. B659, 698 (2008) 

\bibitem{a}D. Amati, M. Ciafaloni and G. Veneziano, Phys. Lett. B 216, 41 (1989)
\bibitem{z1}P.  Dzierzak, J.  Jezierski, P.  Malkiewicz, and W. Piechocki,  Acta Phys. Polon. B41, 717 (2010) 

 \bibitem{skdj}V. Husain, D. Kothawala and S. S. Seahra, Phys. Rev. D 87, 025014 (2013)
\bibitem{d} M. Kober, Phys. Rev. D 82, 085017 (2010)
  \bibitem{1a}   M. Kober, Int. J. Mod. Phys. A 26, 4251 (2011)
  \bibitem{1b} M. Faizal and B. Majumder, Annals Phys. 357,  49  (2015)
  \bibitem{ha}M. Faizal, Int. J. Geom. Meth. Mod. Phys. 12, 1550022 (2015) 
  \bibitem{h}  M. Faizal and S. I. Kruglov, Int. J. Mod. Phys. D 25, 1650013 (2016) 


 


\bibitem{m}  M. Faizal, M. M. Khalil and S. Das,  Eur. Phys. J. C 76,  30 (2016)  
\bibitem{faiz} M. Faizal, A.  F.  Ali and A.  Nassar,   Int. J. Mod. Phys. A 30, 1550183 (2015)

\bibitem{w1}T. D. Lee and G. C. Wick,  Nucl. Phys. B 9, (1969) 
\bibitem{w2}T.  D.  Lee  and  G.  C.  Wick,    Phys. Rev. D 2, 1033 (1970) 
\bibitem{stan} B. Grinstein, D. O'Connell and M. B. Wise, Phys. Rev. D 77, 025012 (2008)
\bibitem{stand} J. R. Espinosa, B.  Grinstein, D.  O'Connell and  M. B. Wise,  	Phys. Rev. D 77, 085002 (2008)
\bibitem{stand1} T. R. Dulaney and M. B. Wise, Phys. Lett. B 658, 230 (2008)
   \end{thebibliography}
\end{document}